\documentclass[namedreferences]{solarphysics}

\usepackage[hyperref,optionalrh]{spr-sola-addons} 
\usepackage{graphicx,geometry}
\usepackage{color}           
\usepackage{breakurl}        

\newcommand{\fact}{\mathcal{F} (\nu)}
\newcommand{\factnt}{\mathcal{F} (\nu,t)}

\newcommand{\mnfact}{\overline{\mathcal{F}}(\nu)}
\newcommand{\Ssun}{S_\odot (\nu,t)}
\newcommand{\Sm}{S^\prime (\theta,\phi,\nu,t)}
\newcommand{\R}{\mathrm{R}_{30} }
\newcommand{\MYhref}[3][blue]{\href{#2}{\color{#1}{#3}}}

\newcommand{\Tb}{$T_{\mathrm B}\ $}

\newcommand{\aap}{    {\it Astron. Astrophys.}}

\newcommand{\apj}{    {\it Astrophys. J.}}
\newcommand{\apjl}{   {\it Astrophys. J. Lett.}}

\newcommand{\pasa}{   {\it Pub. Astron. Soc. Australia}}

\newcommand{\solphys}{{\it Solar Phys.}}

\chardef\us=`\_

\begin{document}

\begin{article}
\begin{opening}

\title{4D Data Cubes from Radio-Interferometric Spectroscopic Snapshot Imaging.\\ {\it Solar Physics}}

\author[addressref={aff1},corref,email={atul@ncra.tifr.res.in}]{\inits{A.}\fnm{Atul}~\lnm{Mohan}\orcid{0000-0002-1571-7931}}
\author[addressref=aff1,email={div@ncra.tifr.res.in}]
{\inits{D.}\fnm{Divya}~\lnm{Oberoi}\orcid{0000-0002-4768-9058}}

\address[id=aff1]{National Centre for Radio Astrophysics - Tata Institute of Fundamental Research, S. P. Pune University Campus, Ganeshkind, Pune 411007, India. }
\runningauthor{Atul M., D. Oberoi}
\runningtitle{4D Radio Data Cubes from Spectroscopic-Snapshot Imaging}

\begin{abstract}
The new generation of low radio-frequency interferometric arrays have enabled the imaging of the solar corona at high spectro-temporal resolutions and sensitivity. In this article we introduce and implement a formalism to generate flux density and brightness temperature ($T_{\mathrm B}$) maps from such images, using independently obtained disc-integrated solar flux density dynamic spectra. These images collectively generate a 4D data cube, with axes spanning angular coordinates ($\theta,\phi$), frequency ($\nu$) and time ($t$). This 4D data cube is the most informative data product which can be generated from interferometric radio data. It will allow us to track solar emission simultaneously in these four dimensions. We also introduce SPatially REsolved Dynamic Spectra (SPREDS), named in analogy to the usual dynamic spectra. For any arbitrary region, ($\theta_i,\phi_j$), in the maps, these 2D projections of the 4D data cube correspond to the dynamic spectrum of emission arising from there. We show examples of these data products using observations from the {\it Murchison Widefield Array} (MWA). These are also the first calibrated solar maps from the MWA.
\end{abstract}
\end{opening}

\section{Introduction}
     \label{Intro} 

It has long been recognised that radio imaging studies with high time and frequency resolution at metre wavelengths are necessary to understanding the physics of solar corona. The ability to study the coronal dynamics in all four dimensions, namely the two angular coordinates, $\theta$ and $\phi$, frequency, $\nu$, and time, $t$, is essential for the detailed investigation of coronal plasma and magnetic field dynamics. The new generation of low radio frequency interferometric arrays have now made this possible by providing snapshot-spectroscopic imaging at sub-second temporal and sub-MHz spectral resolution. Examples of these arrays include the {\it LOw Frequency ARray} (LOFAR; \citealp{vanHaarlem13}) operating in the ranges 10--80 and 120--240 MHz, the {\it Long Wavelength Array} (LWA; \citealp{Kassim2010}) at 10--88 MHz band and the {\it Murchison Widefield Array} (MWA; \citealp{Tingay}) in the 80--300 MHz band. These arrays are already producing new and interesting science results enabled by their novel capabilities ({\it e.g.} \citealp{morosan2015}; \citealp{tunbeltran15}; \citealp{akshay17}). Not only do they provide powerful tools to study long-standing problems like the generation and evolution of various known types of radio bursts, but they also open up new phase space with significant discovery potential.\\
At metre-wavelengths, the bulk of the quiet-Sun solar emission is thermal bremsstrahlung from the million K corona and the non-thermal emissions are predominantly from plasma emission mechanisms. We note that the emission from these plasma emission mechanisms is coherent in nature, implying that even energetically weak events tend to have a high brightness temperature ($T_{\mathrm B}$) associated with them.
So when it comes to looking for signatures of energetically weak features, the metre-wave band has a significant edge over the soft X-ray (SXR) or extreme ultra-violet (EUV) bands, where the observational signature is purely thermal in nature. This makes metre-wave bands a very promising place to look for signatures of weak non-thermal emissions. We note that the spatial resolution at metre-wave bands is much coarser than at EUV and X-ray bands. On the other hand, the low radio-frequency interferometric arrays provide very high resolution in time. Traditionally, the imaging dynamic range (DR) at metre-wave bands has been much lower than that at EUV and X-ray bands. With the availability of new arrays like the MWA, this difference has dropped to a factor of few. All the images used for this work have a DR well beyond 100:1. This work is a part of our effort to build the appropriate tool-set for extracting relevant information from the voluminous data that will be generated by these modern interferometric arrays.\\
Generating well-calibrated flux density and \Tb maps at low radio frequencies is challenging. We build upon recent work by \cite{Div17} who have developed a technique for reliable solar flux calibration, and we extend it to the image domain to make calibrated flux density and \Tb maps. Although this technique is of general applicability, we demonstrate this by applying it to spectroscopic snapshot solar images from the MWA. The collection of these images can be thought of as a four-dimensional (4D) data cube, {\it e.g.} $T_{\mathrm B}(\theta, \phi, \nu, t)$, which encapsulates all the information contained in the radio interferometry data organised along physically meaningful axes. Ideas along these lines have been used by various researchers in the past:  \cite{Div11} applied it to the data from the 32-tile MWA prototype, \citeauthor{morosan2014} (\citeyear{morosan2014}, \citeyear{morosan2015}) have used it for the multi-beam data from LOFAR, and \cite{chen15} for the data from {\it Jansky Very Large Array}. As the use of the new generation of interferometers becomes more widespread, we expect that researchers will find such 4D data products increasingly useful. For ease of use, we also introduce another data product, the SPatially REsolved Dynamic Spectra (SPREDS), which is a 2D projection of the 4D data product, and we show some examples from the MWA.\\
This article is organised as follows: Section 2 provides the formalism for generating the flux density and \Tb maps, the details of the data used and level of activity of the Sun on the day of observations are provided in Section 3, practical details of implementation of the formalism along with an estimate of uncertainties involved are provided in Section 4, the 4D and 2D data cubes (SPREDS) are introduced in Section 5, and Section 6 presents the conclusions.

\section{Formalism}
\label{form}
Making well-calibrated solar flux density or \Tb maps is challenging. With flux densities of even the quiet Sun ranging from a few to many SFUs (1 SFU = $10^4$ Jy), the Sun is much brighter than typical calibrator sources. Most instruments, including the MWA, do not have a linear working range large enough to permit observations of the Sun and the typical calibrator sources. Hence, the solar signals are usually attenuated by as much as a few orders of magnitude. The antenna-based attenuators are often hard to calibrate individually in the field. A procedure to achieve absolute solar flux calibration has recently been developed by \cite{Div17}. Developed in the context of the MWA, it relies upon two important features of the MWA: that it is a reasonably well characterised instrument, and it has many short baselines for which even the Sun remains unresolved. The procedure also assumes that the Galactic and extra-galactic radio emission is described well by the available models. Thus, one can account for all the known contributions to the normalised correlation measured on a short baseline, except for the contribution from the Sun, which can then be estimated. Here we build upon this prescription which provides disc-integrated solar flux density, $S_\odot (\nu,t)$, to generate calibrated flux density and \Tb maps, using:

\begin{equation}\label{eqn1}
S_\odot (\nu,t)=\oiint \limits_{C} S(\theta,\phi,\nu,t)\ d\Omega = \frac{2k\nu^2}{c^2}\oiint \limits_{C} T_\odot (\theta,\phi,\nu,t)\  d\Omega,
\end{equation}
where $S (\theta,\phi,\nu,t)$ and $T_\odot (\theta,\phi,\nu,t)$ are the solar flux density (or brightness) and brightness temperature distribution of the Sun, $k$ is the Boltzmann constant, $c$ is the speed of light and $C$ is the closed contour defining the boundary of the Sun in the image plane.\\ 
Prior to the flux density calibration, the observed flux density in the images, $\Sm$, is in arbitrary units. To relate $\Sm$ to the independently known $\Ssun$, we introduce a quantity, $\factnt$. This quantity inherits its frequency dependence from the response function of the attenuator used for attenuating the solar signal. In the most general case, the attenuator response function can also be a function of time. In practice, the attenuator response is quite stable, especially over the short observing intervals being considered here, which are of the order of minutes. Therefore, we ignore its time dependence and define $\fact$ as  
\begin{equation}\label{facteq}
\Ssun = \fact \oiint \limits_{C} \Sm\ d\Omega.
\end{equation}    
The calibrated flux density maps, $S(\theta,\phi,\nu,t)$, can be arrived at using the scaling relation 
\begin{equation}\label{seq}
S(\theta,\phi,\nu,t) = \fact \Sm,
\end{equation}
where $S(\theta,\phi,\nu,t)$ is in units of ${\mathrm Jy}/\mathrm{\Omega_{\mathrm beam}}$, with $\mathrm{\Omega_{\mathrm beam}}$ the solid angle corresponding to the synthesised clean-beam (referred to simply as the beam hereafter). From these flux density maps, \Tb maps can also be generated in a straightforward manner using  
\begin{eqnarray}\label{tbeq}
T_{\mathrm B}(\theta,\phi,\nu,t)=\frac{1}{\Omega_{\mathrm beam}} \oiint \limits_{\mathrm beam} T_\odot (\theta,\phi,\nu,t)\  d\Omega = \frac{S(\theta,\phi,\nu,t) c^2}{2k\nu^2 \Omega_{\mathrm beam}}.
\end{eqnarray}
\section{Observations and State of the Sun}
	\label{obs}

\begin{figure}    
   \centerline{\hspace*{0.0\textwidth}
               \includegraphics[width=0.32\textwidth,height=1.8in,clip=]{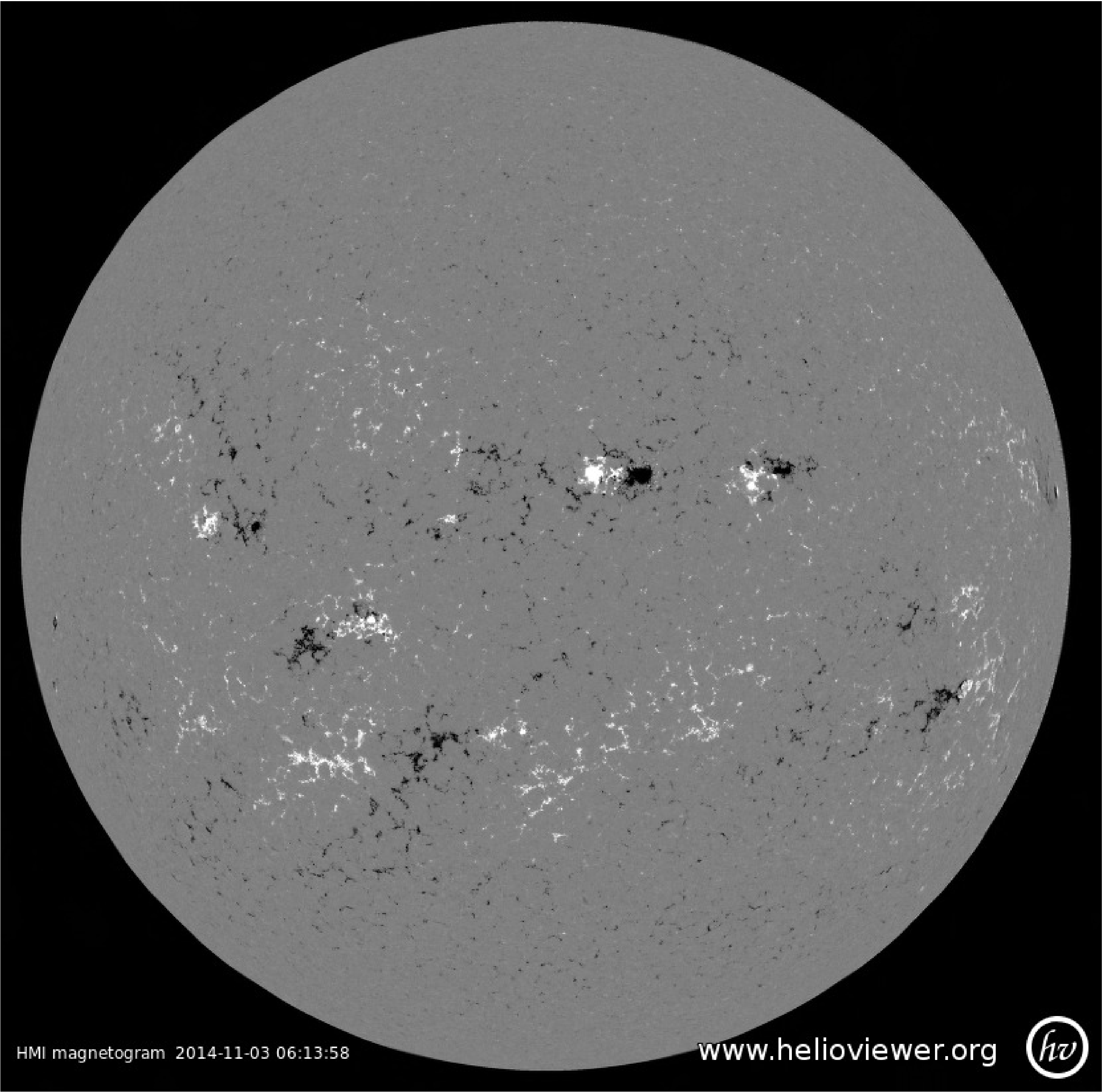}
               \hspace*{-0.01\textwidth}
               \includegraphics[width=0.66\textwidth,height=1.98in,clip=]{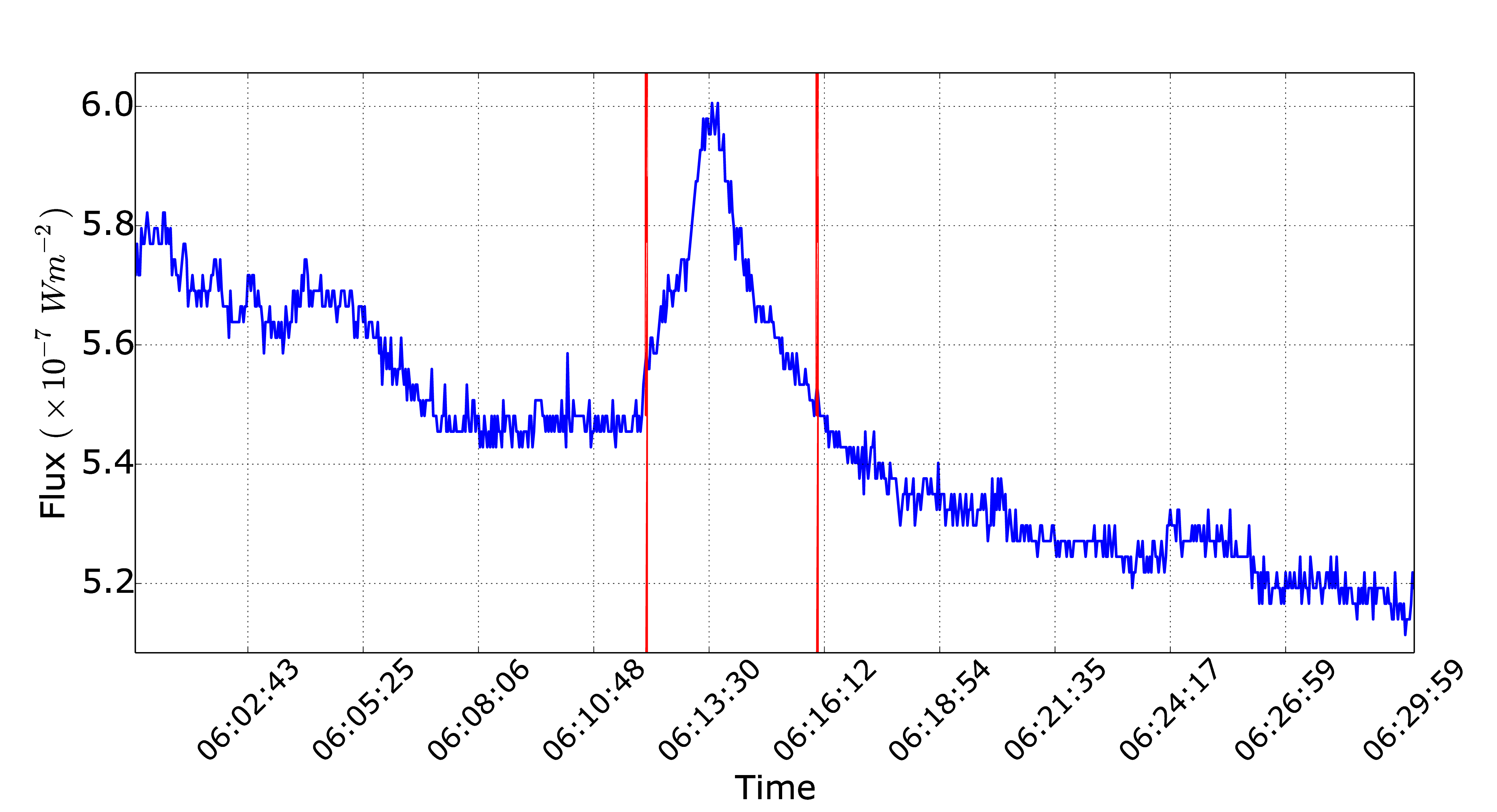}
              }
           
  \caption{{\it Left panel:} HMI magnetogram image of the Sun for the time 06:13:48 UTC, {\it i.e.} around the mid-time of our observations (\MYhref{https://helioviewer.org/}{https://helioviewer.org/}). {\it Right panel: } GOES 1-8 \AA\ light curve. The region between the two vertical lines marks the period of our observation (\MYhref{http://www.lmsal.com/solarsoft/ssw_service}{http://www.lmsal.com/solarsoft}). }
   
  \label{culg}
  \end{figure}

\begin{figure}
\centering
\includegraphics[width=1.00\textwidth,height=0.35\textwidth,clip=]{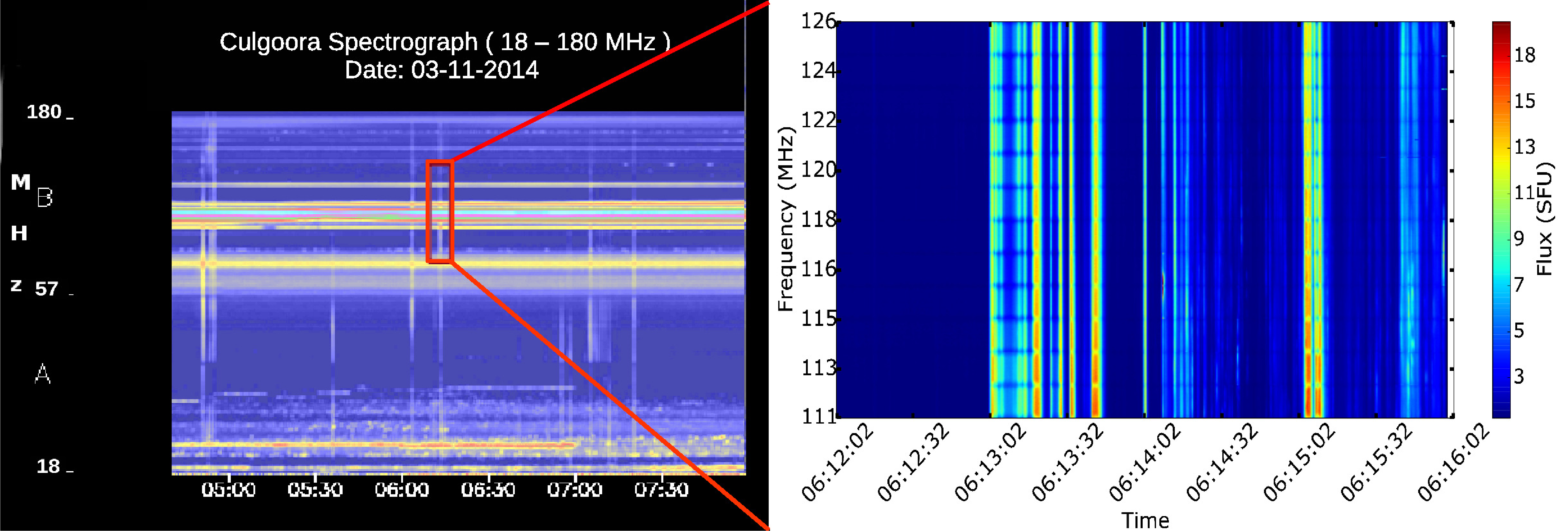}
\vspace{0.0001\textwidth}
\caption{{\it Left panel: } part of the Culgoora dynamic spectrum on 3 November, 2014 (\MYhref{http://www.sws.bom.gov.au/solar}{http://www.sws.bom.gov.au/solar}). The MWA DS during the time period 06:12:02 - 06:16:02 UT in the spectral band from 111 - 126 MHz is shown on the right. The {\it vertical stripes} in the MWA dynamic spectrum are all solar in origin, but, the {\it regular horizontal stripes} are due to instrumental artefacts. The MWA data reveal richer detail in the emission than is seen in Culgoora spectrograph. The MWA DS shown has been flux-calibrated using the prescription by \cite{Div17}.}
\label{DS}
\end{figure}

A precursor to the {\it Square Kilometre Array-Low} (SKA-Low), the MWA is located in Western Australia and operates in the 80--300 MHz band. Its 128 elements are distributed over a $\approx$3 km diameter in a heavily centrally condensed configuration. Each of these elements is an array of 4$\times$4 dual-polarisation dipoles. Details of the MWA design are available in \cite{Lonsdale09} and \cite{Tingay}, the description of the correlator in \cite{ord15}, and the key science targets in \cite{Bowman2013}. The data used here come from MWA observations on 3 November 2014 from 06:12:02-06:16:02 UT, taken under the solar observing proposal G0002. The data presented here have a bandwidth of 15.36 MHz, are centred at 118.4 MHz, and have a time and frequency resolution of 0.5 s and 40 kHz respectively. \\
The Space Weather Prediction Center/National Oceanic and Atmospheric Administration ( SWPC/NOAA) event list for this day reports a total of ten type III radio bursts, eight type IV radio bursts and a C1.4 class flare at 04:50 UT. The NOAA active region summary lists ten active regions on the visible part of the solar disc and the day was characterised as a period of medium activity level\footnote{\MYhref{http://www.solarmonitor.org}{See  http://www.solarmonitor.org}}. We note that no radio event was reported during our observing period. Figure \ref{culg} shows a magnetogram from the {\it Helioseismic and Magnetic Imager} (HMI; \citealp{Scherrer12}) on board the {\it Solar Dynamics Observatory} (SDO; \citealp{Pesnell2012}) and the 1--8 \AA\ light curve from the {\it Geostationary Operational Environment Satellites} (GOES) for a period including that of our observations. The light curve shows a well-defined peak in emission during our observing interval at around 06:13:40 UT \footnote{\MYhref{http://legacy-www.swpc.noaa.gov/ftpdir/}{See  http://legacy-www.swpc.noaa.gov/ftpdir/}}. Although typical of a microflare ($\approx$A5.5 class), this weak feature does not meet the SWPC/NOAA operational definition of a flare. The figure shows that the relative brightness of the soft X-ray background itself was high during that period.
At about the same time, a faint type III feature is seen by the {\it Culgoora Radio Spectrograph} (Figure \ref{DS}, left panel). The right panel of Figure \ref{DS} shows the flux density MWA dynamic spectrum (DS) corresponding to the box marked in its left panel. Owing to its better sensitivity, higher resolution and a very low radio-frequency interference environment, the same feature is seen to have a rather elaborate temporal structure with many short-lived episodes of emission in the MWA data. The vertical stripes arise due to solar emission. In the corresponding radio maps, this strong emission is seen as bright compact sources on the solar disc. The horizontal stripes in the MWA DS arise from instrumental artefacts. A coronal jet is observed to form between 06:12:49 and 06:13:01 UT in all observation bands of the {\it Atmospheric Imaging Assembly} (AIA; \citealp{Lemen2012}), on board SDO, as the result of a prominence reconnection. A detailed study of this jet and the associated type III burst will be the subject of a subsequent paper.  

\section{Implementation}
	\label{tbmap}
	
The format of the correlator output was converted into that of a measurement set suitable for downstream analysis with the MWA specific software named Cotter (\citealp{offringa15}). Solar images were made at a spectral and temporal resolution of 160 kHz and 0.5 s, respectively, using the Common Astronomy Software Applications (CASA; \citealp{casa}) package. The phases of the antenna gains were obtained using observations of a calibrator source, conducted 9 hours prior to the solar observations. The relative amplitude calibration was made using solar observations under the assumption that all tiles are identical. A single cycle of phase-only self-calibration was performed to improve the phase solutions. All of these data, excluding the spectral channels suffering from instrumental artefacts, were imaged for the entire 4 minutes at the aforementioned spectral and temporal resolution, leading to $\approx$23 000 images.\\
As argued in Section \ref{form}, we can justifiably neglect the time dependence of $\factnt$. The corona is a continuous medium, with the electron density steadily dropping with increasing coronal height. Hence the corresponding brightness distribution does not have a sharp edge. This implies that there is no objective criterion for choosing the solar boundary. This subjectivity involved in the choice of the contour defining the boundary of the Sun is the primary source of uncertainty in determining $\fact$ (Equation \ref{facteq}). Next, we describe our approach to quantitatively assess this uncertainty.
	
\subsection{Defining the Solar Boundary}   
In order to assess the impact of the choice of contour defining the boundary of the Sun on the numerical value of $\fact$, we took the following approach. 
The imaging dynamic ranges of the solar images obtained are of the order of a few hundred, and are insufficient to show the solar disc in the presence of type III emission. For the purpose of estimating an appropriate contour to delimit the Sun, we chose to work with images that do not have type III emission and for which the solar disc can be seen clearly. Seven such time slices were identified. For each of them, a set of 48 images was made that spanned the entire band, and it was visually verified that the solar disc could be seen clearly in all of these maps. A set of seven representative images, one from each time slice and also spanning the entire bandwidth of observation, was chosen from these 336 (7$\times$48) images. For each of these images, we defined a small region with a high signal-to-noise ratio (S/N) on the disc of the Sun and examined how the \Tb corresponding to it varied with the choice of the solar boundary contour. While defining regions and boundaries, we chose to work in units of the rms variation , $\sigma$, in the image in a large region far from the Sun. The region in the image with a flux density in the range 30.0 -- 30.1$\sigma$ was defined to be the high-S/N region and is denoted by $\R$. The solar boundary contours were varied from 3$\sigma$ to 14$\sigma$ in steps of 0.5$\sigma$. For the boundary contour, $C$, corresponding to each of these values of $n\sigma$, $\fact$ was computed using Equation \ref{facteq}. It was used to scale the observed brightness distribution to units of SFUs and the corresponding \Tb map was computed using Equation \ref{tbeq}.  
\begin{figure}
\centering
\includegraphics[width=4.6in,height=2.6in]{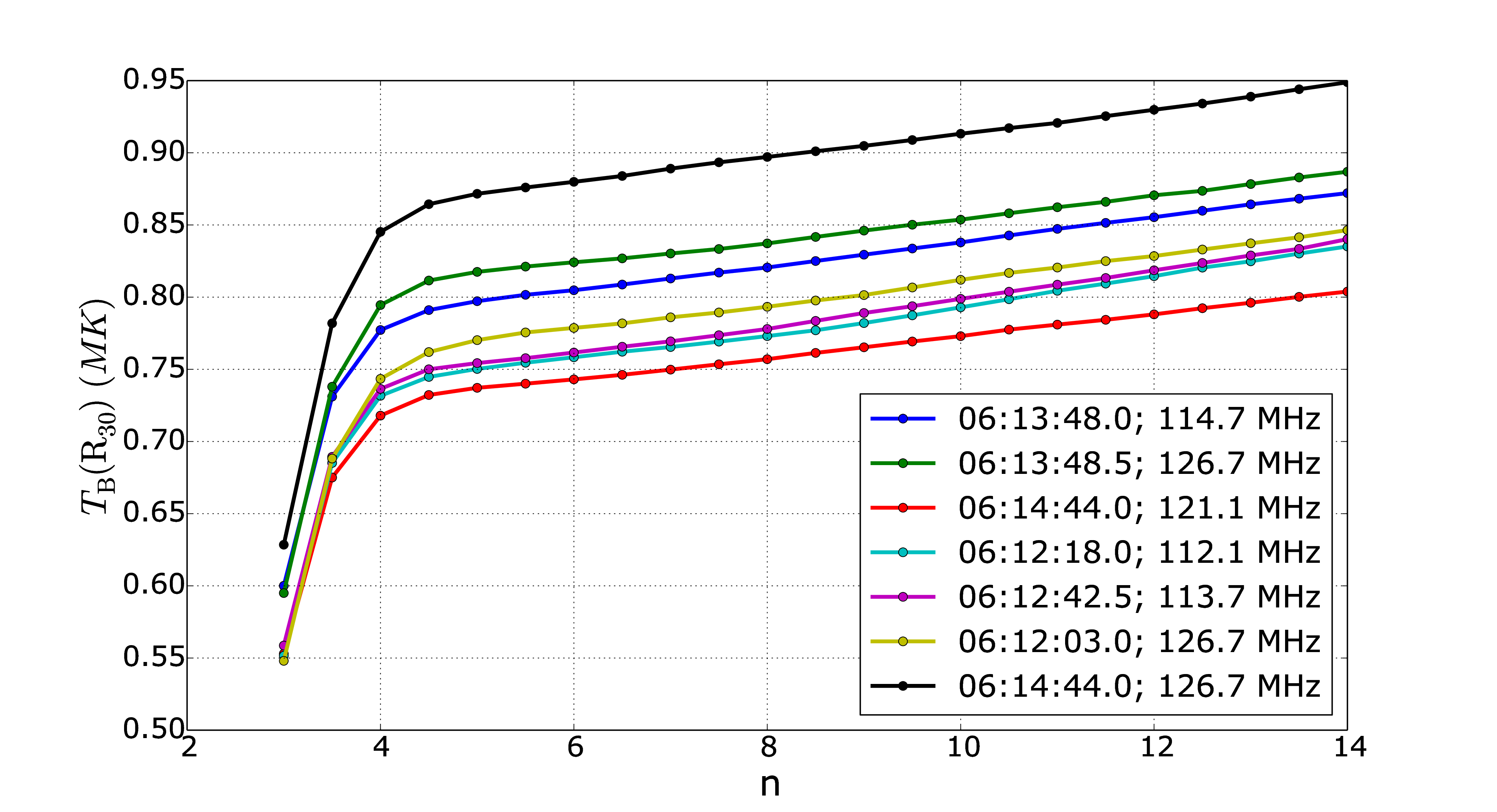}

\caption{Variation of \Tb in a fixed region, $\R$, on the Sun with the choice of the solar boundary. $\R$ is defined as the region in the image with fluxes within $30$ and $30.1\sigma$. The \Tb values obtained for seven different quiet-time images are shown in seven different colours.}
\label{tbVsnsig}
\end{figure}
Figure \ref{tbVsnsig} shows the \Tb estimates for $\R$ as a function of $n$ for each of the 7 time slices. As expected, the estimated \Tb does have a dependence on the choice of the contour chosen that marks the boundary of the Sun. While still in the high-S/N regime this dependence is quite weak. The estimated \Tb values change by less than 10\% as the value of $n$ is varied from 6 to 14. Only when $n$ drops to values below 5 does the estimated \Tb fall steeply. In this work, we use the $6\sigma$ contour to mark the boundary of solar emission.
\subsection{Uncertainty in $\fact$}
\begin{figure}
\centering
\includegraphics[width=5.2in,height=3.5in]{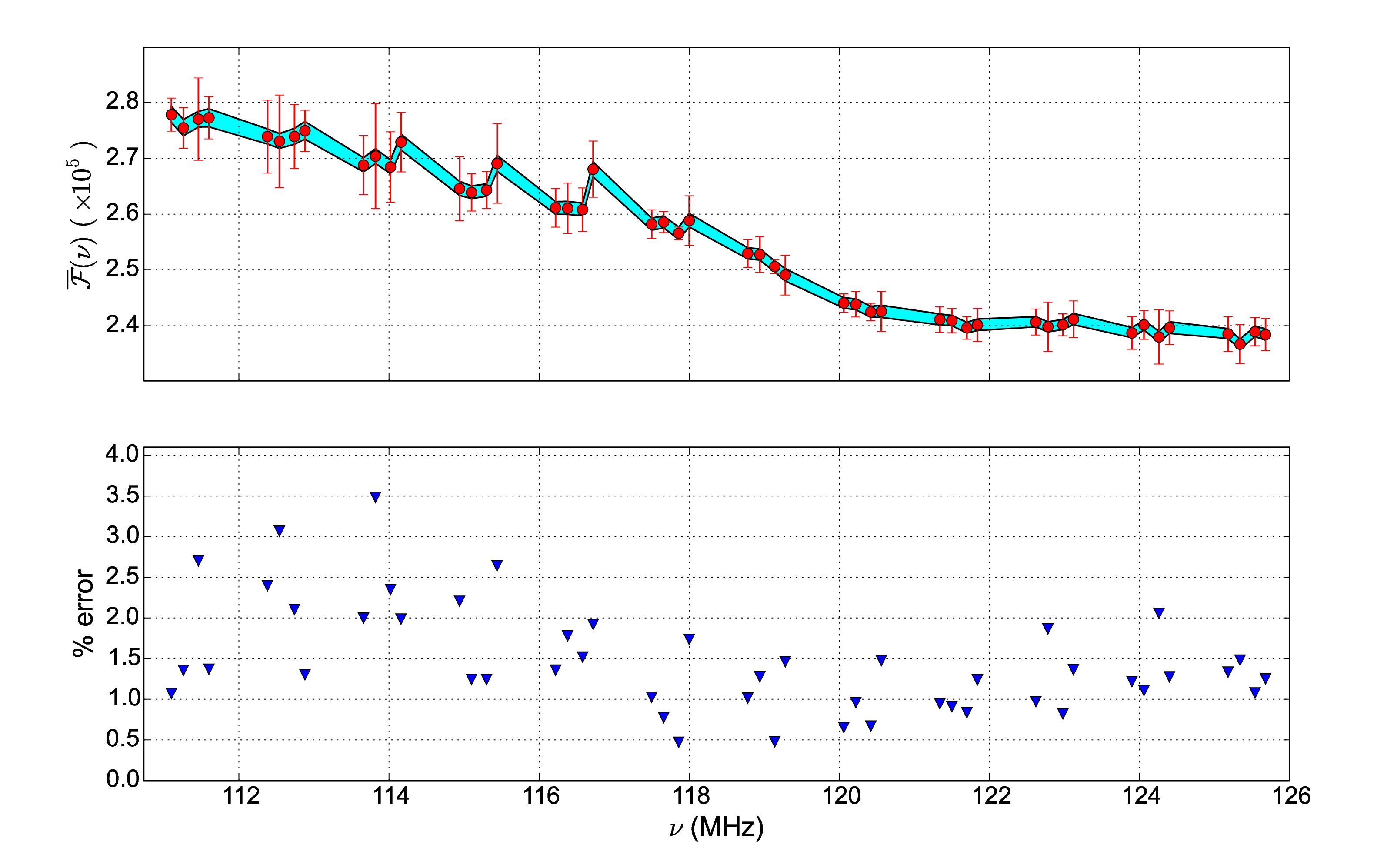}
\caption{{\it Top panel:} $\mnfact$ as a function of $\nu$. The error bars on $\mnfact$ show the rms variation in the individual values of $\fact$ determined at a given $\nu$. {\it Bottom panel: }rms variation in $\fact$ in percentage units. The uncertainty in the estimate of $\fact$ is always found to be $\lesssim 4\%$. The regular data gaps across the spectrum are a consequence of instrumental artefacts. }
\label{mnfact}
\end{figure}

In order to estimate the uncertainty in $F(\nu)$ for our choice of the solar boundary contour ($6\sigma$), we computed $F(\nu)$ for each of the seven chosen time slices across the entire observing band. Figure \ref{mnfact} shows the mean and the rms variation in the value of $F(\nu)$ as a function of frequency. We denote the mean value of $\fact$ by $\mnfact$. The bottom panel plots the percentage error in $F(\nu)$, which is always $\lesssim$ 4\%.\\
The choice of the value of $n$ used to mark the solar boundary contour also leads to an uncertainty in the value of $F(\nu)$. This uncertainty is systematic in nature. This is illustrated by the blue band in the top panel of Figure \ref{mnfact}, which shows the variation in $\mnfact$ as $n$ is varied from 5.5 to 6.5.
As is evident, the systematic error is always smaller than the random component of the uncertainty. 
Moreover, the uncertainty associated with $S_{\odot}(\nu, t)$ can itself exceed 10\% (\citealp{Div17}), so the net uncertainty in the \Tb images is only weakly dependent on the subjective choice made while determining $F(\nu)$.
\section{4D Data Cubes: $S(\theta,\phi,\nu,t)$ and $T_{\mathrm B}(\theta, \phi, \nu, t)$}
\begin{figure}
   \centerline{\hspace*{0.0\textwidth}
               \includegraphics[width=0.35\textwidth,height=1.95in,clip=]{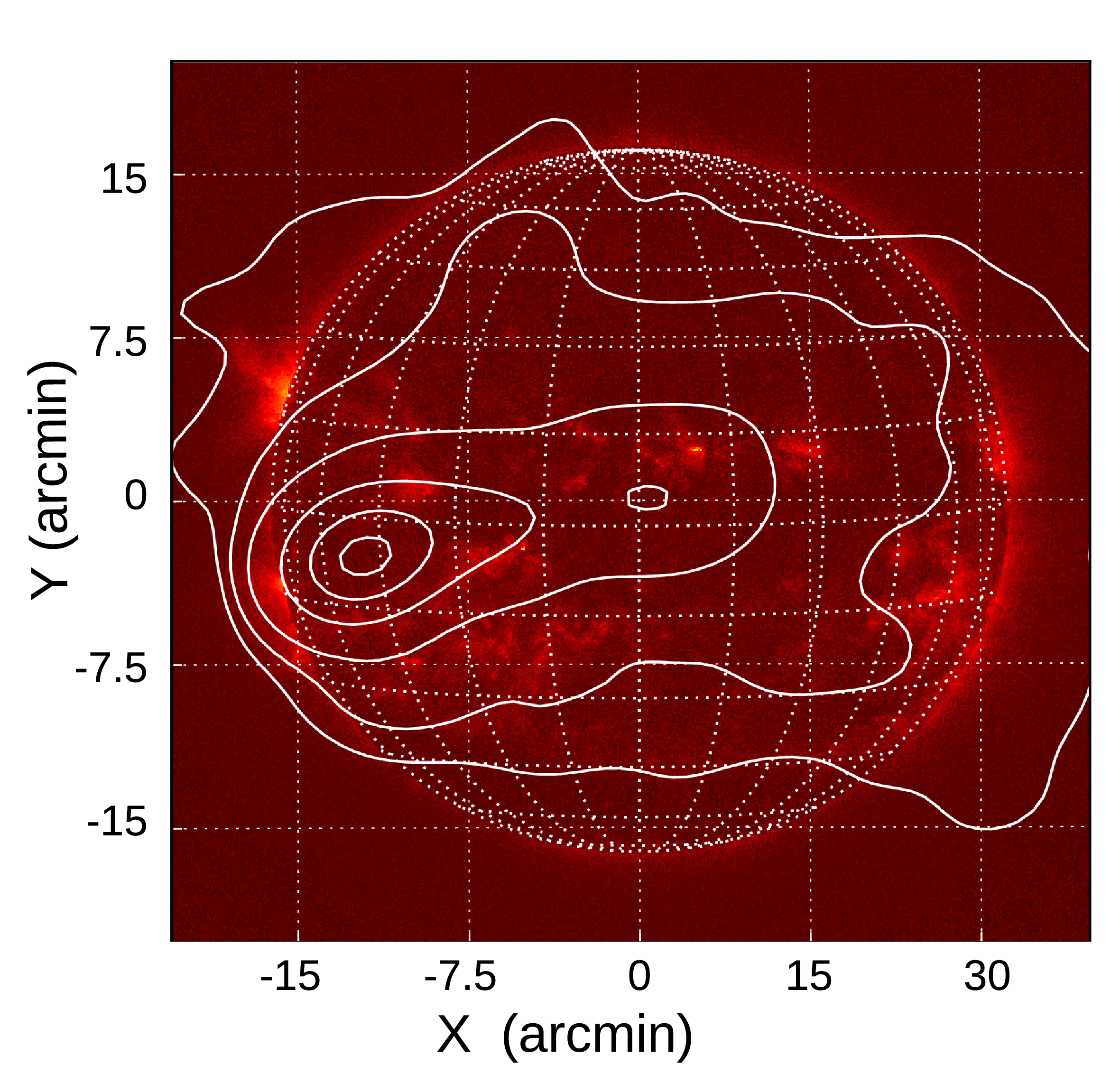}
               \hspace*{-0.008\textwidth}
               \includegraphics[width=0.64\textwidth,height=2in,clip=]{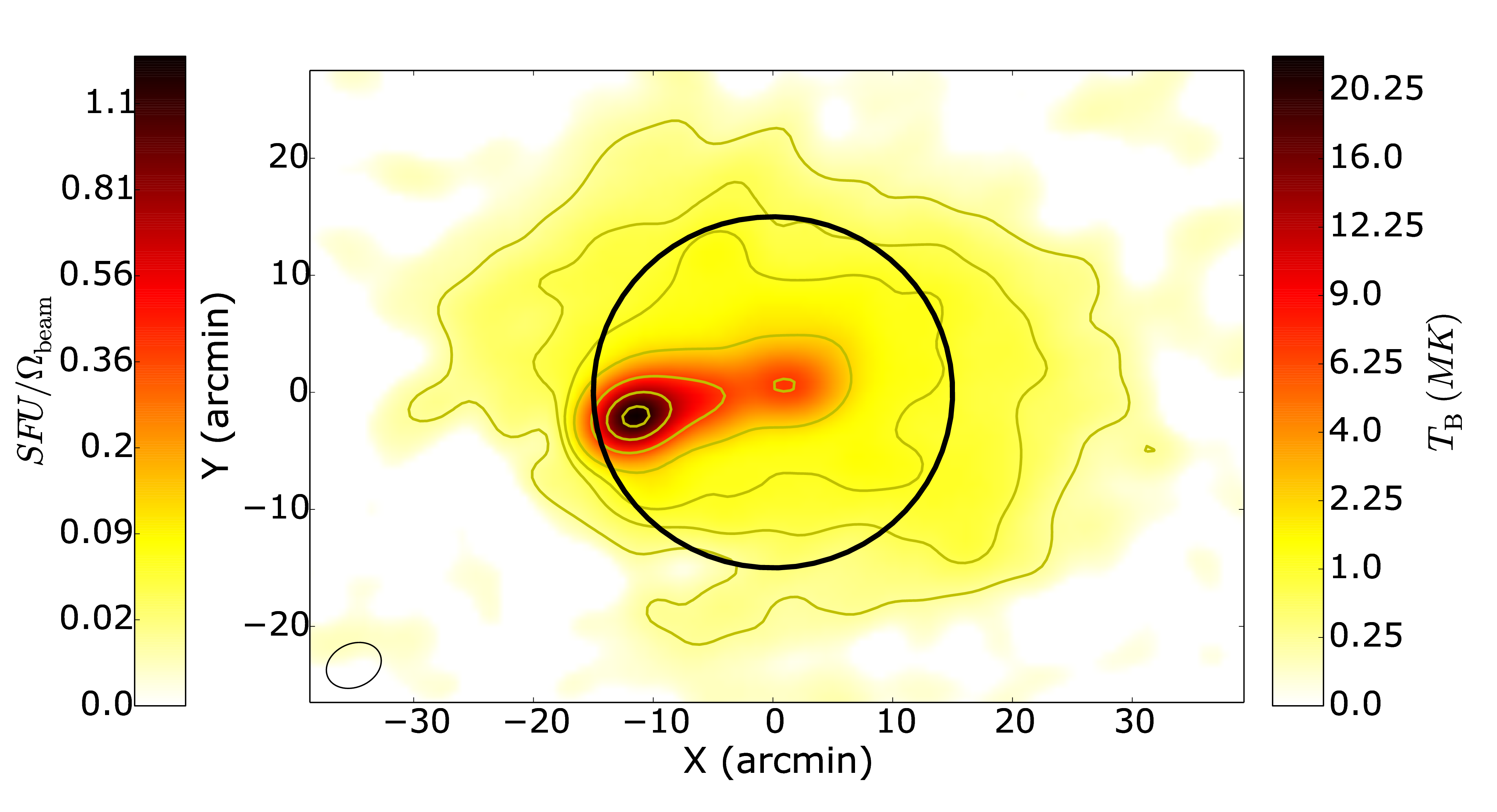}
              }

\caption{{\it Left panel:} AIA 94 \AA\ image with MWA 125.18 MHz contours overlaid. The MWA image at 06:14:06 UT is made with a time and frequency integration of 0.5 s and 160 kHz. The contours marked are at 3\%, 5\%, 10\%, 30\%, 60\% and 90\% of the peak. {\it Right panel: } shows the same MWA image, and the contour levels also include 1\% of the peak. The \Tb scale clearly brings out the non-thermal nature of the brightest emission region as the \Tb value rises by an order of magnitude above the expected coronal temperature of $10^6$ K. The thick circle represents the optical disc and the ellipse in the bottom left corner depicts the synthesised beam.} 
\label{tbS}
\end{figure}

After we computed $F(\nu)$, we could easily convert the uncalibrated solar maps into flux-
density and \Tb maps using Equations \ref{seq} and \ref{tbeq}, respectively. This was done for all $\approx$23 000 maps. These are the first calibrated solar maps from MWA. This collection of maps spanning the time- and frequency-space of the observations represents the most informative data product that can be produced from interferometric data, {\it i.e.} $T_{\mathrm B}(\theta, \phi, \nu, t)$ and $S(\theta,\phi,\nu,t)$. These are 4D data cubes with two axes along the two dimensions in the image plane and the other two along the frequency and time axes, respectively. They allow one to explore the nature of solar emission simultaneously across the time, frequency, and angular coordinate axes. $S(\theta,\phi,\nu,t)$ or $T_{\mathrm B}(\theta, \phi, \nu, t)$ can be thought of as a 2D stack of dynamic spectra, one for each of the independent resolution element (beam) sized regions on the Sun, or as a 2D stack of images, one for each pixel in the DS.\\
$T_{\mathrm B}$ is an intrinsic property of the source, invariant with distance from the source. The relevant solid angle to use for \Tb maps is $\mathrm{\Omega_{\mathrm beam}}$ (Equation \ref{tbeq}).
For unresolved sources the measured \Tb only provides a lower limit to the true $T_{\mathrm B}$. The flux density, on the other hand, is integrated over the solid angle of the source and hence depends on the distance to the source.
For a source that always remains unresolved, the observed \Tb changes with the resolution of the instrument, but its flux density remains constant. For studies of the extended solar emission, \Tb maps are very useful. As stated earlier, the bremsstrahlung from the MK corona forms the thermal emission at these frequencies. This implies than any emission with a \Tb higher than the coronal temperature must be powered by  non-thermal processes, providing a robust criterion for identifying the presence of non-thermal coronal emission. The right panel of Figure \ref{tbS} shows an example radio image with a colour scale in units of both flux density and $T_{\mathrm B}$. We note that for convenience, the values assigned to pixels in radio flux density maps are in units of Jy beam$^{-1}$, making the pixel values independent of the size of the pixels in the map. This also implies that flux density and \Tb maps only differ by a scale factor. In this work, we rely primarily on using \Tb images. In the left panel of the same figure we overlay the MWA 125.18 MHz contours on an AIA 94 \AA\ image for comparison.

The active regions in our images are larger than the beam even during burst times, so we use \Tb maps and $T_{\mathrm B}(\theta, \phi, \nu, t)$ data cube for the rest of the discussion. 
\begin{figure}
\centering
\hspace*{-0.00\textwidth}
\includegraphics[height=2.5in,width=5.74in]{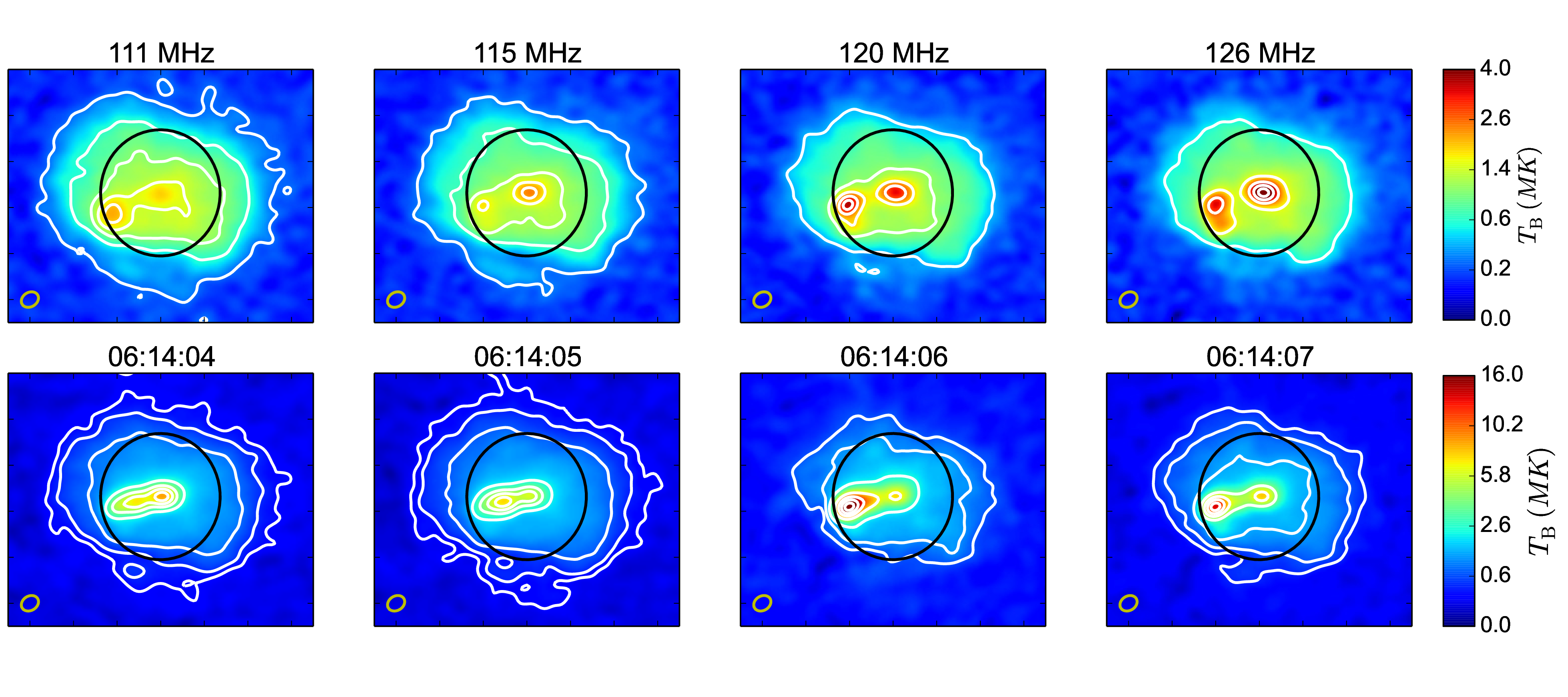}
\caption{Collage of \Tb maps of the Sun at different times and frequencies. The {\it top row} shows images at 06:12:50 UT for different frequencies spanning the observing band. The contour levels are marked at 10\%, 30\%, 50\%, 70\% and 90\% of the peak. The {\it bottom row} shows solar images at 125.18 MHz for different times, one second apart. The contour levels in this row are marked at 3\%, 5\%, 10\%, 30\%, 50\%, 70\% and 90\% of the peak. The {\it thick circle} represents the optical solar disc. The color bars in the images are in square root scale.  The {\it ellipse} in the bottom left corner of every image depicts the synthesised beam.}
\label{tbcollage}
\end{figure}
Figure \ref{tbcollage} shows a collage of a selection of \Tb maps. The top row shows a set of maps from the same time slice but at different frequencies, spanning the observing bandwidth. The bottom row shows a set of maps from the same spectral slice but 1s apart. These images correspond to relatively quiet time periods of the Sun. They serve to illustrate the rapid changes across the time, frequency and angular coordinate axes, and the ability of $T_{\mathrm B}(\theta, \phi, \nu, t)$ to capture this wealth of information.

\subsection{SPatially REsolved Dynamic Spectrum (SPREDS)}
	\label{SPREDS}	
While $T_{\mathrm B}(\theta, \phi, \nu, t)$ is indeed the most comprehensively informative data product, it is extremely voluminous. If generated at the finest MWA resolution over its 30.72 MHz bandwidth, each of these 4D data cubes will comprise 0.37$\times 10^6$ radio images of the Sun every 4 minutes. With instruments with a much wider bandwidth on the horizon, such as the SKA-Low, assuming similar spectral resolution, the number of maps will grow by about an order of magnitude (\citealp{ska-low}). Even with our spectral averaging and dropping the spectral channels affected by instrumental artefacts, we still obtain about $\approx$23 000 images every 4 minutes. So, in spite of its very rich information content, it is rather hard for the human mind to assimilate this information. While these are the natural axes for the description of the data, in all likelihood, there exist axes along which the same information can be expressed in a much more compact manner. The $T_{\mathrm B}(\theta, \phi, \nu, t)$ data will provide excellent applications for machine learning and artificial intelligence based algorithms and techniques to distil its information content into compact forms that are easier to assimilate and interpret.\\ 
In the interim however, once $T_{\mathrm B}(\theta, \phi, \nu, t)$ has been created, it is very instructive to examine its slices across different axes. For instance, a very informative data product can be obtained by freezing the two angular dimensions of $T_{\mathrm B}(\theta, \phi, \nu, t)$ to specific values ($\theta_i, \phi_j$). We denote this data product by ${T_{\mathrm B}}_{\theta_i, \phi_j} (\nu, t)$ and name it the SPatially REsolved Dynamic Spectrum (SPREDS), in analogy with the usual definition of DS. The SPREDS is simply the DS of emission originating from a resolution element, centred at a specific location ($\theta_i, \phi_j$) in the solar maps. An enormous amount of knowledge in solar astronomy has been gained by studying the DS from radio spectrographs, in spite of the obvious limitation that they do not provide any information about the location or shape of the emission region. This success stems primarily from the facts that owing to the coherent nature of plasma emission mechanisms, the active solar radio emissions outshine the thermal solar emission by a very large margin (a few to many orders of magnitude); the emission is often expected to come from a compact region, implying that details of morphology are not particularly important or informative; and the comparatively bright features that have been the usual subjects of these investigations are infrequent enough that it is plausible to assume that only one of them occurs on the Sun at any given time.  As we extend our studies to features with extended morphologies and weaker emissions using the nascent high dynamic-range imaging capabilities of the new generation of instruments, simultaneous studies of features and dynamics at vastly varying strengths will become possible. In these situations SPREDS will form a very effective tool.      
\begin{figure}
\centering
\includegraphics[width=14cm, height=11.5cm]{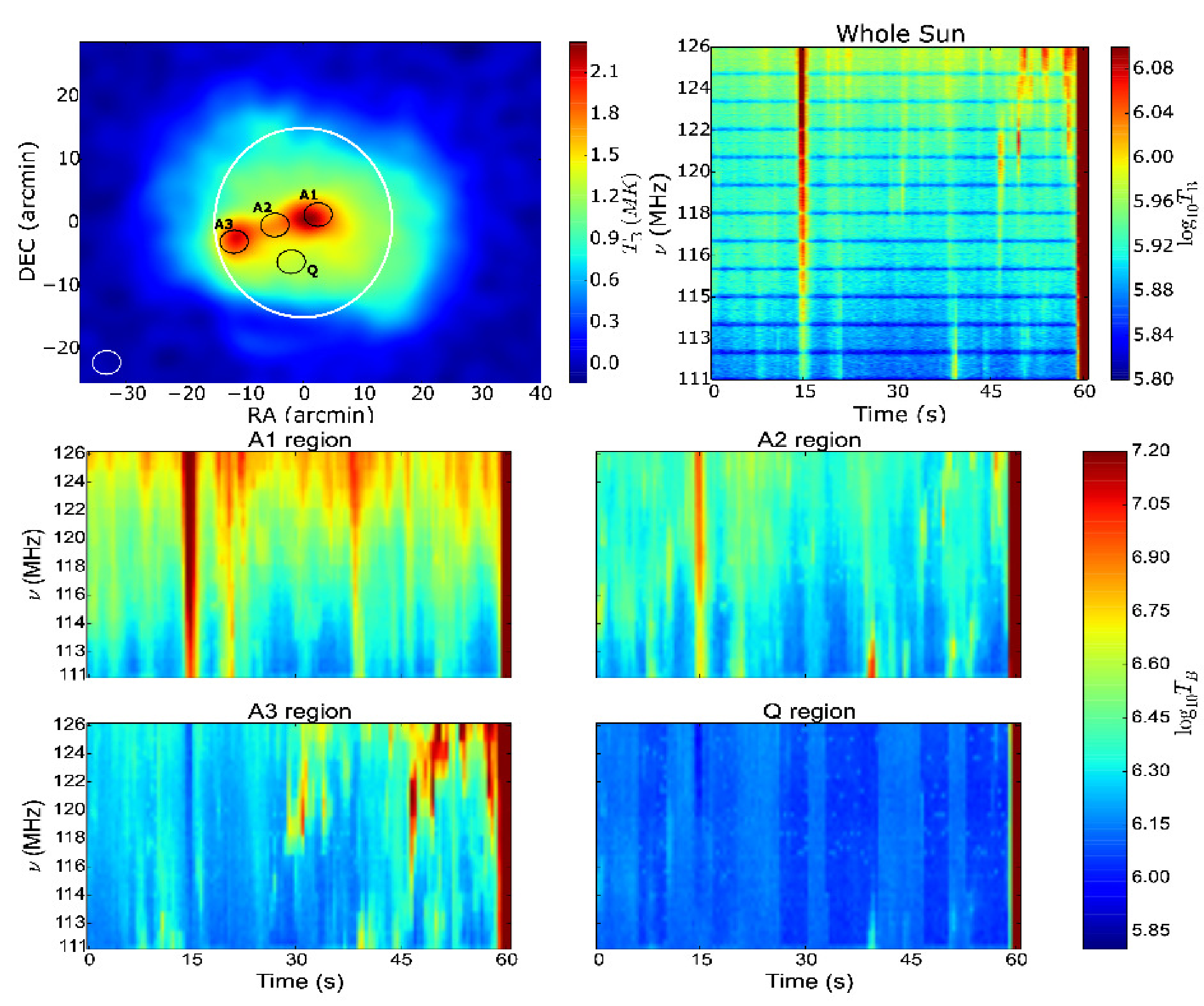}
  \caption{{\it Top left panel: } example image of the Sun with four beam-sized regions marked, namely A1, A2, A3 and Q. The regions marked by A are active at various times during the observations and Q is a relatively quiet region of the radio Sun. A2 is the region where the type III radio burst source appears.  {\it Top right panel: }mean \Tb DS of the entire Sun for the first minute of the observation. The second and third rows show the SPREDS plots for all the four regions for the same period. These plots clearly show the differences between the emissions arising from these locations. The bright feature at $\approx$60 s in all the DS plots is concurrent with the coronal jet event. The colour scale is logarithmic and has been saturated to highlight the features of the radio Sun during relatively quiet observation times.}
\label{localDS}
\end{figure}
Figure \ref{localDS} is an illustration of this. The top left panel shows an example radio map of the Sun with various regions of interest marked on it. Regions A1-A3 show enhanced emission at different times throughout the observations and Q marks a quiet region that shows comparatively weak emission and little variability. All the type III emissions are seen to come from the A2 region. The bottom two rows show the SPREDS corresponding to these four different regions on the Sun during a period of relatively low solar activity, {\it i.e.} the first $\approx$1 minute of the observations. To capture the emission features spanning a large range, a logarithmic colour scale has been chosen. The scale is saturated to highlight the spatial variability of the fine emission features in the SPREDS. We note that for the usual DS, the observed value of \Tb is an average over the entire solar disc. For compact emission this leads to a significant reduction in the observed $T_{\mathrm B}$. On using SPREDS this spatial averaging is limited to the size of the instrumental resolution (beam), significantly improving both the contrast and the fidelity of the measurement. Clearly, the different weak emission features seen in the solar mean \Tb DS in the top right panel are contributed to by different regions on the Sun. For example, consider the group of faint features between $\approx$45 and 60 s seen in the mean \Tb DS. They are caused by emission from the A3 region. The strong feature at $\approx$15 s in the mean \Tb DS is primarily due to A1. We do see a correlated brightening at A2 as well which is due to the flux leakage from A1 to A2. Although in this instance the correlation arises from flux leakage, SPREDS provides a very useful tool for identifying and exploring such correlations, {\it e.g.} in the presence of long-range magnetic fields and/or arcades. The \Tb at the region Q is of the order of $10^6$ K throughout, which is the expected coronal temperature at the heights we are probing. The bright red edge to all the SPREDS and the mean \Tb DS is due to the bright radio burst at A2 with a $T_{\mathrm B}$ $\approx 10^9$ K. This feature is so bright that the enhancement in emission due to it even at the locations of A1, A3 and Q is sufficient to saturate the chosen colour scale. SPREDS thus lets us isolate interesting emissions from different regions of the Sun. \\
\section{Conclusion}
	\label{concl}
Using independently obtained disc-integrated solar flux densities, we have developed a prescription to generate flux density and \Tb maps of the Sun from low radio-frequency interferometric data. We demonstrated this procedure using an example dataset from the MWA. A complete set of such radio images can be organised in the form of 4D data cubes of flux density or $T_{\mathrm B}$. These data structures provide a convenient construct that encompasses exhaustive information about solar emission across the axes of angular coordinates, frequency and time. We also introduced a particularly useful 2D projection of this 4D data cube, SPREDS, the dynamic spectrum corresponding to a specific region on the Sun. We showed examples of these that were generated using data from the MWA. These are the first calibrated solar maps from MWA. \\
With spectroscopic imaging from the new generation of instruments now becoming available, these would be the natural data products to work with and we expect them to find increasing use. As we try to understand the Sun in greater detail, these tools will become indispensable because they can distinguish the emission along the four physically meaningful axes. Along with access to new information, these 4D data cubes also bring their own challenges. The more manageable ones relate to the heavy computational and storage burden for
generating and storing these images. The larger challenge will perhaps lie in finding ways to
identify and extract the interesting information from this enormous number of images in an
intelligible and accessible form and to use this information to make progress in solar radio
physics. We believe this will be fertile ground for applications of techniques from the fields
of machine-learning and computer-aided discovery.

\begin{acks}
This scientific work makes use of the {\it Murchison Radio-astronomy Observatory}, operated by the Commonwealth Scientific and Industrial Research Organisation (CSIRO). We acknowledge the Wajarri Yamatji people as the traditional owners of the Observatory site. Support for the operation of the MWA is provided by the Australian Government through the National Collaborative Research Infrastructure Strategy (NCRIS), under a contract to Curtin University administered by Astronomy Australia Limited. We acknowledge the Pawsey Supercomputing Centre which is supported by the Western Australian and Australian Governments. The data used here are also a courtesy of the SDO, WIND and GOES science teams. We are grateful to the other members of the group, namely S. Mondal and R. Sharma for their suggestions at various stages of the work and help in generating some plots. We thank the developers of Python 2.7\footnote{\MYhref{https://docs.python.org/2/index.html}{See https://docs.python.org/2/index.html}} and the various associated packages especially Matplotlib \footnote{\MYhref{http://matplotlib.org/}{See http://matplotlib.org/}}, Astropy \footnote{\MYhref{http://docs.astropy.org/en/stable/}{See http://docs.astropy.org/en/stable/}} and NumPy\footnote{ \MYhref{https://docs.scipy.org/doc/}{See https://docs.scipy.org/doc/}}.\\

\noindent \textbf{Disclosure of Potential Conflicts of Interest}$\quad$ The authors declare that they have no conflicts of interest.
\end{acks}
    

\end{article} 

\end{document}